\documentclass[showpacs,amsmath,amssymb,amsfonts,a4paper,aps,prd]{revtex4}
\usepackage{graphicx}

\newcommand{\f}[2]{\frac{#1}{#2}}

\def\be{\begin{equation}}
\def\ee{\end{equation}}
\def\bea{\begin{eqnarray}}
\def\eea{\end{eqnarray}}
\def\bl{\begin{align}}
\def\el{\end{align}}

\begin{document}

\title{Matter may matter\footnote{Essay written for the Gravity Research
Foundation
2014 Awards for Essays on Gravitation}}
\author{Zahra Haghani$^1$}
\email{z.haghani@du.ac.ir}
\author{Tiberiu Harko$^2$}
\email{t.harko@ucl.ac.uk}
\author{Hamid Reza Sepangi$^1$}
\email{hr-sepangi@sbu.ac.ir}
\author{Shahab Shahidi$^1$}
\email{s.shahidi@du.ac.ir}
\affiliation{$^1$School of Physics, Damghan University, Damghan, Iran}
 \affiliation{$^2$Department of Mathematics, University College London, Gower
Street,
London, WC1E 6BT, United Kingdom}
\affiliation{$^3$Department of Physics, Shahid Beheshti University, G. C.,
Evin,Tehran 19839, Iran}

\begin{abstract}

We propose a gravitational theory in which the effective Lagrangian of the
gravitational field is given by an arbitrary function of the Ricci scalar, the
trace of the matter energy-momentum tensor, and the contraction of the Ricci
tensor with the matter energy-momentum tensor.  The matter energy-momentum
tensor is generally not conserved, thus leading to the appearance of an
extra-force, acting on  massive particles in a gravitational field. 
The stability conditions of the theory with respect to local perturbations are also obtained. The cosmological implications
of the theory are investigated, representing an exponential solution. Hence a Ricci tensor - energy-momentum tensor coupling may
explain the recent acceleration of the Universe, without resorting to the
mysterious dark energy.
\end{abstract}


\date{\today}

\maketitle


The recent observational advances in cosmology have opened  new windows for the
understanding of the basic properties of the Universe and of the gravitational
interaction that dominates its dynamics and evolution. The Planck satellite data
of the 2.7 degree Cosmic Microwave Background  full sky survey \cite{1,2} have
generally confirmed the standard $\Lambda $Cold Dark Matter cosmological
paradigm. The recent measurement of the tensor modes from
large angle CMB B-mode polarisation by BICEP2 \cite{3}, implying a
tensor-to-scalar ratio $r = 0.2 ^{+0.07}_{-0.05}$, has provided a very
convincing evidence for the inflationary scenario, since  the generation of
gravity-wave fluctuations is a generic prediction of the de Sitter expansion. However,  the BICEP2 result is  in tension with Planck limits on
standard inflationary models \cite{Mar}. 
The observation of the accelerated expansion of the Universe
\cite{Perlmutter,Riess1,Riess2} has raised the fundamental issue of the cause of
this acceleration, which is usually attributed to a mysterious and yet not
directly detected dominant component of the Universe, called dark energy
\cite{Copeland:2006wr}.

 The numerous observational successes of cosmology require sound explanations
and justifications and the need of putting cosmology, and more generally
gravity, on a sound theoretical basis is felt as ever before. However, presently
there is no convincing theoretical model, supported by observational evidence,
which could clearly explain the nature of dark energy.  Moreover, the recent
accelerated expansion of the Universe,  the galaxy rotation curves  and the virial mass discrepancy at the
galactic cluster level \cite{Str}, usually
explained by postulating the existence of another mysterious and yet undetected
component of the Universe, the so-called dark matter,  suggest that the
standard general relativistic gravitational field equations, based on classic
Einstein-Hilbert
action $S=\int{\left(R/2+L_m\right)\sqrt{-g}d^4x}$, where $R$ is the scalar
curvature, and $L_m$ is the matter Lagrangian density,  in which matter is
minimally coupled to the geometry, cannot give an appropriate quantitative
description of the Universe at large scales, beyond the boundary of the Solar
System. From a cosmological point of view, these phenomena require the {\it ad hoc} introduction in the  energy-momentum tensor, in addition to ordinary baryonic matter, of the dark matter and dark energy components.

In going beyond the Einstein-Hilbert action the first step  was to generalize the
geometric part of the action.  An extension
of the Einstein-Hilbert action, with an arbitrary function of
the scalar invariant, $f(R)$, has been extensively explored in the literature
\cite{rev}. Such a modification of the gravitational action  accounts for the
late acceleration of the Universe and may also provide  a geometric explanation
for dark matter, which can be understood as a manifestation of  gravity itself \cite{Boehm}.
The fascination for extra-dimensions, going back to the unified
field theory of Kaluza and Klein, has led to the development of the brane world
models \cite{Randall1, Randall2, Maartens1}. In brane world models extra-dimensional gravitational effects dominate at
high energies, but significant new effects that can
successfully explain both dark energy and dark matter, do appear at low energies.  Quadratic Lagrangians,
constructed from second order curvature invariants such as $R^{2}$, $R_{\mu \nu
}R^{\mu\nu }$, $R_{\alpha \beta \mu \nu }R^{\alpha \beta \mu \nu }$,
$\varepsilon ^{\alpha \beta \mu \nu }R_{\alpha \beta \gamma \delta
}R_{\mu \nu }^{\gamma \delta }$, $C_{\alpha \beta \mu \nu}C^{\alpha \beta \mu
\nu}$, etc., have also been considered as candidates for a more general
gravitational action \cite{Lobo:2008sg}.

Most of the modifications of the Einstein-Hilbert Lagrangian did concentrate
only on the geometric part of the action and assumed that the matter term does
play a subordinate and passive role, guaranteed by its minimal coupling to
geometry. However, a theoretical principle forbidding a more general coupling
between matter and geometry does not exist {\it a priori}. If such a coupling is
allowed, theoretical gravitational models with many interesting features can be
constructed.  By following the early works in \cite{Bertolami:2007gv} and
\cite{Harko:2008qz}, a maximal extension of the Einstein-Hilbert action of the
form $S=\int d^{4}x \sqrt{-g}f\left(R,L_m\right)$ was considered in
\cite{Harko:2010mv}. An extension of  $f(R,L_\mathrm{m})$
gravity is the $f(R,T)$ model \cite{Harko:2011kv}, with the corresponding
action given by
$$S=\frac{1}{16\pi}\int
d^{4}x\sqrt{-g}f\left(R,T\right)+\int{d^{4}x\sqrt{-g}L_\mathrm{m}},$$
where $T$ is the trace of the energy-momentum tensor of the
matter, $T_{\mu \nu}$. The dependence on $T$ may be due to the presence of  quantum effects (conformal anomaly), or of some
exotic imperfect fluids.

However,  $f\left(R,L_m\right)$ or $f(R,T)$ types of gravitational Lagrangians
are not the most general Lagrangians with
non-minimal  matter - geometry coupling. For example, one may generalize the
above modified theories of gravity by introducing a term $R_{\mu\nu}T^{\mu\nu}$
in the Lagrangian \cite{fRT,Od}. Examples of such couplings can also be found in the
Einstein-Born-Infeld theories \cite{deser}, when one expands the square root in
the Lagrangian.
When $T=0$, which is the case of electromagnetic radiation, the field equations of $f(R,T)$ gravity reduce to those of $f(R)$
gravity. However, considering the presence of the $R_{\mu\nu}T^{\mu\nu}$
coupling term still entails a non-minimal coupling of geometry to the electromagnetic field.

We therefore propose to describe the gravitational field by the following
action, taking into  account a  coupling between the energy-momentum tensor of
ordinary matter, $T_{\mu \nu}$, and the Ricci curvature tensor $R_{\mu \nu}$,
\be\label{eq200}
S=\f{1}{16\pi G}\int d^4x\sqrt{-g}f\left(R,T,R_{\mu\nu}T^{\mu\nu}\right)+\int
d^4x\sqrt{-g}L_m.
\ee
By varying the action Eq.~(\ref{eq200}) with respect to the metric we
obtain the gravitational field equations as \cite{fRT}
\be\label{eq204}
G_{\mu\nu}+\Lambda_{eff}g_{\mu\nu}=8\pi G_{eff}T_{\mu\nu}+T^{eff}_{\mu\nu},
\ee
where the effective gravitational coupling $G_{eff}$, the
effective cosmological constant $\Lambda _{eff}$, and the effective
energy-momentum tensor $T^{eff}_{\mu\nu}$ are
\be\label{eq204-1}
G_{eff}=\f{G+\f{1}{8\pi}\big(f_T+\f{1}{2}f_{RT}R-\f{1}{2}\Box
f_{RT}\big)}{f_R-f_{RT}L_m},
\ee
\be
\Lambda_{\tiny{eff}}=\f{2\Box
f_R+Rf_R-f+2f_TL_m+\nabla_\alpha\nabla_\beta(f_{RT}T^{\alpha\beta})}{2(f_R-f_{RT
}L_m)},
\ee
and
\bl\label{eq204-2}
T^{eff}_{\mu\nu}&=\f{1}{f_R-f_{RT}L_m}\Bigg\{\nabla_\mu\nabla_\nu
f_R-\nabla_\alpha f_{RT}\nabla^\alpha T_{\mu\nu}
   -\f{1}{2}f_{RT}\Box T_{\mu\nu}-
2f_{RT}R_{\alpha(\mu}T_{\nu)}^{~\alpha}\nonumber\\
&+\nabla_\alpha\nabla_{(\mu}\left[T^\alpha_{
~\nu)}f_{RT}\right]
   +2\left(f_Tg^{\alpha\beta}+f_{RT}R^{\alpha\beta}\right)
\f{\partial^2
L_m}{\partial g^{\mu\nu}\partial g^{\alpha\beta}}\Bigg\},
\end{align}
respectively. The index of $f$ denotes the derivative, and $RT$ stands for $R_{\mu\nu}T^{\mu\nu}$. In general $G_{eff}$ and $\Lambda_{eff}$ are not constants, and depend on the specific model considered.

The equation of motion for a massive test particle with the matter Lagrangian
$L_m=p$ takes the form
\be\label{eq404}
\f{d^2x^\lambda}{ds^2}+\Gamma^\lambda_{~\mu\nu}u^\mu u^\nu=f^\lambda,
\ee
with the extra force acting on massive test particles  given by
\bea\label{eq405}
f^\lambda&=&\f{1}{\rho+p}\Bigg[\left(f_T+Rf_{RT}\right)\nabla_\nu\rho-
\left(1+3f_T\right)\nabla_\nu p
-(\rho+p)f_{RT}R^{\sigma\rho}\left(\nabla_\nu
h_{\sigma\rho}-2\nabla_\rho h_{\sigma\nu}\right)\nonumber\\
&&-f_{RT}
R_{\sigma\rho}h^{\sigma\rho}\nabla_\nu\left(\rho+p\right)\Bigg]
\f{h^{\lambda\nu}}{1+2f_T+Rf_{RT}}, \qquad\qquad h^{\alpha \beta }=g^{\alpha \beta }+u^{\alpha }u^{\beta
}.
\eea
The extra force does not vanish even with the Lagrangian $L_m=p$. In the
Newtonian limit the generalized Poisson equation is obtained as
\bea\label{poison}
\nabla^2\phi=\f{1}{2(f_R-2\rho f_{RT})}\bigg[8\pi G\rho+3\nabla^2 f_R-3\rho f_T
& -&2f+ \nabla(3f_R+\rho f_{RT}) \cdot \nabla \phi\bigg],
\eea
where $\rho $ is the energy density of the matter. The generalized Poisson
equation contains the gradient of the Newtonian potential $\phi$. In the same limit, the total acceleration of a massive system,
$\vec{a}$, is given by
\be\label{eq418}
\vec{a}=\vec{a}_N+\vec{a}_E,
\ee
where $\vec{a}_N=-\nabla \phi$ is the Newtonian acceleration, and the
supplementary acceleration, induced by the geometry-matter coupling, is
\be\label{eq419}
\vec{a}_E(\rho)=-\nabla U(\rho)=\f{F_0}{\rho_0}\vec{\nabla}
\rho,
\ee
where $\rho _0$ is the background density and
\be\label{eq413}
F_0=\left.\f{f_T+f_{RT}(R-R_{\alpha\beta}h^{\alpha\beta})}{1+2f_T+Rf_{RT}}
\right|_{\rho =\rho _0}.
\ee

The viability of the theory can be studied by examining  its stability with respect to local perturbations. In pure $f(R)$ gravity, if the function
$f(R)$ satisfies the condition $f''(R) < 0$, a fatal instability, the 
“Dolgov-Kawasaki” instability,  develops on time scales of the order of 
$10^{−26}$ seconds \cite{DK}. In the present
case,  the condition of the stability with respect to the local perturbations
can be formulated as \be
f_{RR}\left(R_0\right)- \left(\rho _0-T_0/2\right)f_{RT,R}\left(R_0\right)\geq
0,
\ee
where $R_0$ is the background Ricci scalar.

Interesting cosmological consequences of the model can be obtained once the
explicit functional form of the function $f$ is chosen. Example of specific
choices for $f$ are $f=R+\alpha R_{\mu\nu}T^{\mu\nu}$, $f=R+\alpha
R_{\mu\nu}T^{\mu\nu}+\beta\sqrt{T}$ and $f=R(1++\alpha R_{\mu\nu}T^{\mu\nu})$,
where $\alpha,~\beta ={\rm constant}$, respectively \cite{fRT}. For  $f=R\left(1+\alpha
R_{\mu\nu}T^{\mu\nu}\right)$, the cosmological gravitational field equations for
a flat Friedmann-Robertson-Walker type geometry are given by
\begin{align}\label{co6}
-3 H^2+\kappa  \rho +\alpha  \bigg(18 H   \ddot{H}\rho+18 H \dot{H} \dot{\rho }
+54 H^2  \dot{H}\rho -9  \dot{H}^2\rho +27 H^3 \dot{\rho } +27H^4 \rho \bigg)=0,
   \end{align}
   and
   \bea\label{co7}
  && -2 \dot{H}-3 H^2+\alpha  \Bigg(6 \dddot{H} \rho +12 \ddot{H} \dot{\rho }+36
H  \ddot{H}\rho
     +6 \dot{H} \ddot{\rho } +  54 H \dot{H}
   \dot{\rho }+48 H^2  \dot{H}\rho +15  \dot{H}^2\rho +9 H^2 \ddot{\rho }+
   30 H^3 \dot{\rho }- 9 H^4 \rho \Bigg)=0,\nonumber\\
   \eea
   respectively, where $H$ is the Hubble function, and $\kappa=8\pi G$. The terms proportional to
$\alpha $ in the generalized Friedmann equations (\ref{co6}) and (\ref{co7})
play the role of an effective supplementary density and pressure, which  may be
responsible for the late time acceleration of the Universe. A de Sitter type
solution of the field equations with $H=H_0={\rm constant}$ and
$a=a_0\exp\left(H_0t\right)$ does exist if the matter density varies as
   \bea
  \rho (t)&=& e^{-\frac{1}{6} H_0 \left(t-t_0\right)} 
\Bigg\{\frac{\sqrt{\alpha}  H_0 \left(H_0 \rho _0+6
   \rho _{01}\right) }{\sqrt{
   145 \alpha  H_0^4+4 \kappa }}
  \sinh \left[\frac{ \sqrt{ 145
   \alpha  H_0^4+4 \kappa }}{6 \sqrt{\alpha } H_0}\left(t-t_0\right)\right]
     +\rho _0 \cosh
   \left[\frac{ \sqrt{145 \alpha  H_0^4+4 \kappa }}{6 \sqrt{\alpha
   } H_0}\left(t-t_0\right)\right]\Bigg\},\nonumber\\
   \eea
where $\rho _0=\rho \left(t_0\right)$ and $\rho
_{01}=\dot{\rho }\left(t_0\right)$, respectively.
In order that the ordinary matter density decays exponentially, $\alpha $ must
satisfy the constraint $\alpha <-\kappa /36H_0^4$. The ultra-high energy density
regime of this model, corresponding to the condition $p=\rho $, has similar
properties for the $p=0$ case, that is, it also admits a de Sitter phase.

 In conclusion, we have proposed a gravitational model with an arbitrary coupling between the energy-momentum
and the Ricci tensors.   An extra force  is always present even in
the case $L_m=p$, and causes a deviation from geodesic paths. The extra force could explain the properties of the galactic rotation curves
without resorting to the dark matter hypothesis.
 The supplementary acceleration is proportional to the matter density
gradient, tending to zero for constant density self-gravitating systems. A
similar dependence on the gradient of the Newtonian gravitational potential also
appears in the generalized Poisson equation.
The cosmological implications of the  theory are also promising, with the
gravitational field equations admitting  de Sitter type solutions.
Matter-geometry coupling may therefore be responsible for  late time
acceleration of the Universe, and matter itself may play a more fundamental role in the gravitational dynamics that usually assumed.  




\begin{thebibliography}{99}

\bibitem{1} P. A. R. Ade et al., Planck 2013 results. I, arXiv: 1303.5062
[astro-ph) (2013).

\bibitem{2} P. A. R. Ade et al., Planck 2013 results. XVI, arXiv: 1303.5076
[astro-ph] (2013).

\bibitem{3} P. A. R Ade et al. [ BICEP2 Collaboration],
arXiv:1403.3985 [astro-ph.CO] (2014).

\bibitem{Mar} C. Bonvin, R. Durrer, and R. Maartens, arXiv:1403.6768 (2014).

\bibitem{Perlmutter}
S.~Perlmutter {\it et al.},   Astrophys.\ J.\  {\bf 517}, 565
(1999).

\bibitem{Riess1} A.~G.~Riess {\it et al.},   Astron.\ J.\  {\bf 116},
1009 (1998).

\bibitem{Riess2}
A.~G.~Riess {\it et al.},   Astrophys.\ J.\  {\bf 607}, 665 (2004).

\bibitem{Copeland:2006wr}
  E.~J.~Copeland, M.~Sami and S.~Tsujikawa, Int.\ J.\ Mod.\ Phys.\ D {\bf 15}, 1753 (2006).

  \bibitem{Str} L. E. Strigari, Physics Reports {\bf 531}, 1 (2013).

  \bibitem{rev} A. De Felice and S. Tsujikawa, Living Rev. Rel. {\bf 13}, 3
(2010); T. P. Sotiriou and V. Faraoni, Rev. Mod. Phys. {\bf 82}, 451 (2010); S.
Nojiri and S. D. Odintsov,  Phys. Rept. {\bf 505}, 59 (2011);
F.~S.~N.~Lobo,
  arXiv:0807.1640 [gr-qc]; T. Clifton, P. G. Ferreira, A. Padilla and C.
Skordis, Phys. Rep. {\bf 513}, 1 (2012).
  
  \bibitem{Boehm} C. G. Boehmer, T. Harko, and F. S. N. Lobo, Astropart. Phys. {\bf 29}, 386 (2008).

  \bibitem{Randall1}
L. Randall L and R. Sundrum,  Phys. Rev. Lett. {\bf 83}, 3370 (1999)
[arXiv:hep-ph/9905221].

\bibitem{Randall2}
L. Randall and R. Sundrum, 
Phys. Rev. Lett. {\bf 83}, 4690 (1999).

\bibitem{Maartens1}
R. Maartens,  Living Rev. Rel. {\bf 7}, 7
(2004).

\bibitem{Lobo:2008sg}
  F.~S.~N.~Lobo,
  ``The Dark side of gravity: Modified theories of gravity,''
  arXiv:0807.1640 [gr-qc].


\bibitem{Bertolami:2007gv}
  O.~Bertolami, C.~G.~Boehmer, T.~Harko and F.~S.~N.~Lobo,
    Phys.\ Rev.\ D {\bf 75}, 104016 (2007).

\bibitem{Harko:2008qz}
  T.~Harko, Phys.\ Lett.\ B {\bf 669}, 376 (2008).

\bibitem{Harko:2010mv}
  T.~Harko and F.~S.~N.~Lobo, Eur.\ Phys.\ J.\ C {\bf 70}, 373 (2010).

\bibitem{Harko:2011kv}
  T.~Harko, F.~S.~N.~Lobo, S.~'i.~Nojiri and S.~D.~Odintsov,
   Phys.\ Rev.\ D {\bf 84}, 024020 (2011).

\bibitem{fRT} Z. Haghani, T. Harko, F. S. N. Lobo, H. R. Sepangi, and S.
Shahidi,  Phys. Rev. D {\bf 88}, 044023 (2013).

  \bibitem{Od} S. D. Odintsov and D. S${\rm \acute{a}}$ez-G\'{o}mez, Phys. Lett.
B {\bf 725}, 437 (2013).

   \bibitem{deser} S. Deser and G. W. Gibbons, Class. Quantum Grav. 15, L35
(1998).

  \bibitem{DK} A. D. Dolgov and M. Kawasaki, Phys. Lett. {\bf B 573}, 1 (2003).

\end{thebibliography}
\end{document}